\documentclass[apj]{emulateapj}

\usepackage{color}
\begin{document}

\title{Follow-Up Observations of PTFO~8-8695: A 3 MYr Old T-Tauri Star\\
Hosting a Jupiter-mass Planetary Candidate}

\author{David R.~Ciardi\altaffilmark{1},
Julian C.~van Eyken\altaffilmark{1,2,3},
Jason W.~Barnes\altaffilmark{4},
Charles A.~Beichman\altaffilmark{1},
Sean J.~Carey\altaffilmark{5},
Christopher J.~Crockett\altaffilmark{6},
Jason Eastman\altaffilmark{3,7},
Christopher M.~Johns-Krull\altaffilmark{8},
Steve B.~Howell\altaffilmark{9},
Stephen R.~Kane\altaffilmark{10},
Jacob N.~Mclane\altaffilmark{11},
Peter Plavchan\altaffilmark{12},
L.~Prato\altaffilmark{11},
John Stauffer\altaffilmark{5},
Gerard T.~van Belle\altaffilmark{11},
Kaspar von Braun\altaffilmark{11}
}
                             
\altaffiltext{1}{NASA Exoplanet Science Institute/Caltech Pasadena, CA USA}
\altaffiltext{2}{University of California, Santa Barbara/LCOGT Santa Barbara, 
CA,USA}
\altaffiltext{3}{Las Cumbres Observatory Global Telescope Network, Goleta, CA 
USA}
\altaffiltext{4}{University of Idaho, Moscow, ID, USA}
\altaffiltext{5}{Spitzer Science Center/Caltech Pasadena, CA, USA}
\altaffiltext{6}{Science News, Washington D.C., USA}
\altaffiltext{7}{Harvard-Smithsonian Center for Astrophysics, Cambridge, MA, 
USA}
\altaffiltext{8}{Rice University, Houston, TX, USA}
\altaffiltext{9}{NASA Ames Research Center, Mountain View, CA, USA}
\altaffiltext{10}{San Francisco State University, San Francisco, CA, USA}
\altaffiltext{11}{Lowell Observatory, Flagstaff, AZ, USA}
\altaffiltext{12}{Missouri State University, Springfield, MO, USA}

\email{ciardi@ipac.caltech.edu}




\begin{abstract}

We present Spitzer 4.5\micron\ light curve observations, Keck NIRSPEC radial 
velocity observations, and LCOGT optical light curve observations of 
PTFO~8-8695, which may host a Jupiter-sized planet in a very short orbital 
period (0.45 days).  Previous work by \citet{vaneyken12} and \citet{barnes13} 
predicts that the stellar rotation axis and the planetary orbital plane should 
precess with a period of $300 - 600$ days.  As a consequence, the observed 
transits should change shape and depth, disappear, and reappear with the 
precession. Our observations indicate the long-term presence of the transit 
events ($>3$ years), and that the transits indeed do change depth, disappear and 
reappear.  The Spitzer observations and the NIRSPEC radial velocity observations 
(with contemporaneous LCOGT optical light curve data) are consistent with the 
predicted transit times and depths for the $M_\star = 0.34\ M_\odot$ precession 
model and demonstrate the disappearance of the transits.  An LCOGT optical light 
curve shows that the transits do reappear approximately 1 year later.  The 
observed transits occur at the times predicted by a straight-forward propagation 
of the transit ephemeris. The precession model correctly predicts the depth and 
time of the Spitzer transit and the lack of a transit at the time of the NIRSPEC 
radial velocity observations. However, the precession model predicts the return 
of the transits approximately 1 month later than observed by LCOGT. Overall, the 
data are suggestive that the planetary interpretation of the observed transit 
events may indeed be correct, but the precession model and data are currently 
insufficient to confirm firmly the planetary status of PTFO~8-8695b. 

\end{abstract}


\section{Introduction}\label{intro-sec}

With the discoveries fueled by the Kepler Mission \citep{borucki10}, there are 
now more than 1800 confirmed or validated planets \citep[e.g.,][]{batalha11, 
rowe14}.  Kepler has increased our knowledge of the diversity of planets and 
planetary systems, and the sheer number of planets discovered by Kepler, coupled 
with on-going discoveries from other transit programs, radial velocity surveys, 
direct imaging efforts, and microlensing campaigns 
\citep[e.g.,][]{gillon14,wittenmyer14,ki12,koshimoto14}, have spawned a 
realization that nature yields planetary systems in a variety of configurations 
\citep[e.g.,][]{wf14, fabrycky14, ciardi13, steffen13}.  To understand the 
processes that shape the planetary systems that we observe today, it is crucial 
to observe the planets during the period of formation and evolution.

Evidence has been found for extra-solar rings around a 16 Myr-old pre-main 
sequence star in the Upper Centaurus-Lupus subgroup of Sco-Cen, possibly 
indicative of early planet formation around a T Tauri star \citep{mamajek12}.  
However, discovering and observing planets in the earliest stages of formation 
and evolution requires observing stars that are only $<5-10$ Myr old, as this is 
the timescale over which the planet forming disks are depleted 
\citep[e.g.,][]{haisch01,hillenbrand08}, although recent work may indicate that 
the disks may last as long as $<20$ Myr \citep{pfalzner14}.  But stars at this 
age are notoriously active and spotted, making them photometrically and 
spectroscopically variable and making planet discovery at these young ages 
difficult \citep[e.g.,][]{miller08,crockett12,cody14}.

The Palomar Transient Factory (PTF) Orion survey, conducted in the winters of 
2009 and 2010 \citep{vaneyken11, vaneyken12}, was an attempt to address this 
deficit by searching for young transiting hot Jupiters in the 25 Ori group, a 
small association of T-Tauri stars identified by \citet{briceno05,briceno07}. In 
\citet{vaneyken12}, we reported the discovery of a promising young planetary 
candidate orbiting the known weak-lined T-Tauri star PTFO~8-8695 with an age 
previously estimated of $\approx 3$~Myr \citep{briceno05}. Superimposed on top 
of notable quasi-periodic variability, the star showed regular transit events 
with a period of just $0.45$ days. But unlike TW~Hya, which is a classical T 
Tauri star with active accretion \citep{rk83} and has periodic $\sim 2$\% dips 
likely caused by the disk occulting a hot spot \citep{siwak14}, PTFO~8-8695, 
being a WTTS, has no infrared excess and likely no disk to occult or veil the 
star.  Finally, no evidence was found for a stellar companion. 
\begin{deluxetable*}{ccccccc}
\tablecolumns{7}
\tablewidth{6in}
\tablecaption{Spitzer Light Curve Data (sample)\label{spitzerlc-tab}}
\tablehead{
\colhead{BJD} &
\colhead{Phase} & 
\colhead{Flux} &
\colhead{Uncertainty} &
\colhead{Flux Model} &
\colhead{Flux$-$Model} &
\colhead{Transit Model}
}
\startdata
2456045.70197 & -0.0014 &  0.9912959 & 0.0032780 & 0.9981057 & -0.0068097 & 
-0.0058880 \\
2456045.70233 & -0.0006 &  0.9932295 & 0.0033324 & 0.9980942 & -0.0048646 & 
-0.0058900 \\
2456045.70269 &  0.0002 &  0.9914524 & 0.0037372 & 0.9980825 & -0.0066302 & 
-0.0058890 \\
2456045.70305 &  0.0010 &  0.9927249 & 0.0039651 & 0.9980709 & -0.0053461 & 
-0.0058860
\enddata

\end{deluxetable*}

The observed optical radial velocity variations (Keck HIRES and HET HRS) were 
found to be out of phase with the transit events and were not of sufficient 
amplitude to be caused by a stellar companion, arguing that the observed 
transits were not the result of an eclipsing binary \citep{vaneyken12}. This is 
in contrast to V471 Tau which shows similar depth eclipses, but also 150 km/s 
amplitude radial velocity variations in phase with the eclipses caused by a 
stellar mass white dwarf orbiting a K-dwarf \citep{kaminski07}. The radial 
velocity variations observed in the optical for PTFO~8-8695 are likely caused by 
the rotation of the spotted stellar surface \citep{prato08,huelamo08,huerta08}. 
The spot induced observed radial velocity variations in the optical ($\sim$ 2 
km/s semi-amplitude) yield a candidate companion mass upper limit of $M\sin i < 
4.8\ M_{Jup}$.  Coupled with the transit modeling and the constraints on the 
line of sight inclination \citep[$i\approx60^\circ$,][]{vaneyken12}, the upper 
limit of the companion mass was found to be $M_p\ \lesssim 5\ M_{Jup}$ -- well 
within the planetary mass regime.

The transit light curves did show shape changes between the two years in which 
the PTF Orion observations were made. \citet{barnes13} showed that these effects 
could result from planetary transits of a fast-rotating, oblate, and 
significantly gravitationally darkened star, which is consistent with the short 
rotation period and high projected rotational velocity of PTFO~8-8685 
\citep[$v\sin i \approx 80$ km/s,][]{vaneyken12}. Being non-spherical, and 
darker at the equator than at the poles \citep{vonzeipel24}, oblique transits of 
such a rapidly rotating star can show unusual and asymmetric transit shapes 
\citep{barnes09}. Furthermore, if the planet orbit is oblique to the stellar 
rotation axis, a torque is exerted on the orbit by the stellar oblateness, 
leading to precession of the orbital nodes. 

Orbital precession has been noted previously in other systems and its effect on 
observed planetary transits - particularly in circumbinary systems 
\citep[e.g.,][]{ll13,mt14}. Recently, such behavior has previously been noted in 
the Kepler~13b planet system \citep{szabo11,szabo12,szabo14,barnes11}, and with 
the circumbinary planet, Kepler~413b \citep{kostov14}. These are predicted to 
precess on timescales of decades. But \citet{vaneyken12} and \citet{barnes13} 
showed that, for PTFO~8-8695, the precession period was on the order of a $300 - 
600$ days -- a timescale accessible to observation.

\citet{barnes13} were able to simultaneously fit both years of the PTF Orion 
data set using a fully consistent model which included gravitational darkening, 
oblateness of the host star, and an analytical model of the precession of the 
system. The fits yielded a planet mass $M_p\sim 3\ M_{Jup}$, and predicted rapid 
precession on approximately year-long timescales. The fits implied a very high 
obliquity for the orbit ($\approx 70^\circ$), resulting in a prediction that the 
transits should change depth, shape, and \emph{disappear} for periods of a few 
months to a year as the precession brings the planet's orbit out of the line of 
sight to the star. The limited timespan of the data and the uncertainty in the 
stellar mass and radius left an unresolved degeneracy in the solution. More 
observations were needed to test the veracity of the model and better determine 
the system parameters.  In particular, observations were needed to test that the 
transits changed shape, disappeared, and reappeared in a predictable manner. 

Towards this end, we obtained follow-up observations of PTFO~8-8695 which 
included Spitzer photometry to confirm the transit events in the infrared; Keck 
NIRSPEC infrared radial velocity measurements to measure the radial velocity 
signature of the planet and the Rossiter-McLaughlin effect \citep{rossiter24, 
mclaughlin24, queloz00, gw07} as the planet transited the star, hinted at in the 
original optical radial velocity data \citep{vaneyken12}; optical photometric 
monitoring with the Las Cumbres Observatory Global telescope Network (LCOGT) to 
support the radial velocity measurements and to better establish the transit 
parameters and orbit. 

The Spitzer data, the Keck NIRSPEC data, and early optical LCOGT photometry data 
were acquired before the realization that the transits could disappear 
completely, but these data, in the end,  provided evidence for the changing and 
disappearing transits.  Further optical LCOGT observations were obtained with 
the realization that the system could precess and that the transits could 
disappear. The optical data show the return of the transits at the time 
predicted by propagating the transit mid-point ephemeris, but approximately 1 
month prior to the prediction of the precession model. Given the complexity of 
the system as described in \citet{vaneyken12} and \citet{barnes13}, we regard 
the partial matching of the observations to the model as an indication that the 
planetary interpretation of the nature of the source of the transits is still 
viable, but we are unable to confirm the planet with these data and the current 
models. 

\begin{figure*}[ht]
     \centering
     \includegraphics[angle=90,scale=0.6,keepaspectratio=true]{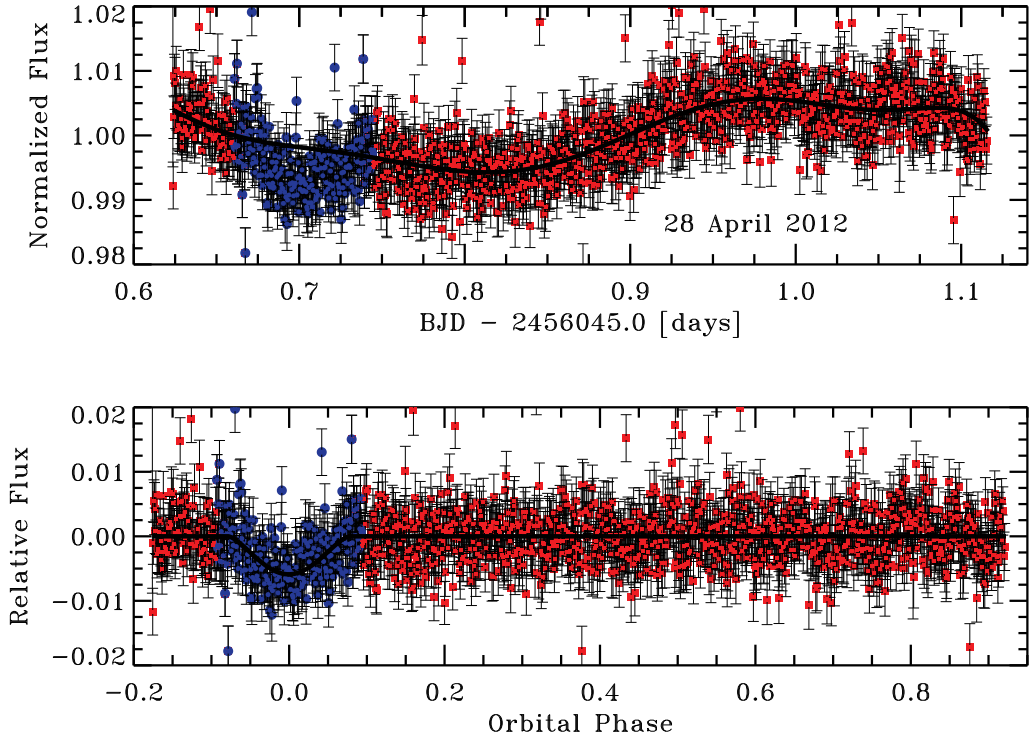}
     \figcaption{\emph{Top:} April 2012 Spitzer 4.5\micron\ light curve plotted 
as a function of time. The blue circles indicate the data points predicted to 
be within the transit window from the PTF ephemeris \citep{vaneyken12}; the red 
squares are outside the transit window and were used to fit an $8^{th}$-order 
polynomial (black line) to model the stellar variability. \emph{Bottom:} The 
Spitzer 4.5\micron\ light curve with the polynomial fit subtracted to remove the 
stellar variability is plotted as function of orbital phase. The solid black 
line represents the transit fit to the light curve and the data have been phased 
on this new ephemeris (see Section~\ref{ephm-sub}).\label{spitzer-fig}} 
\end{figure*}

\begin{figure*}[ht]
     \centering
    \includegraphics[angle=90,scale=0.6,keepaspectratio=true]{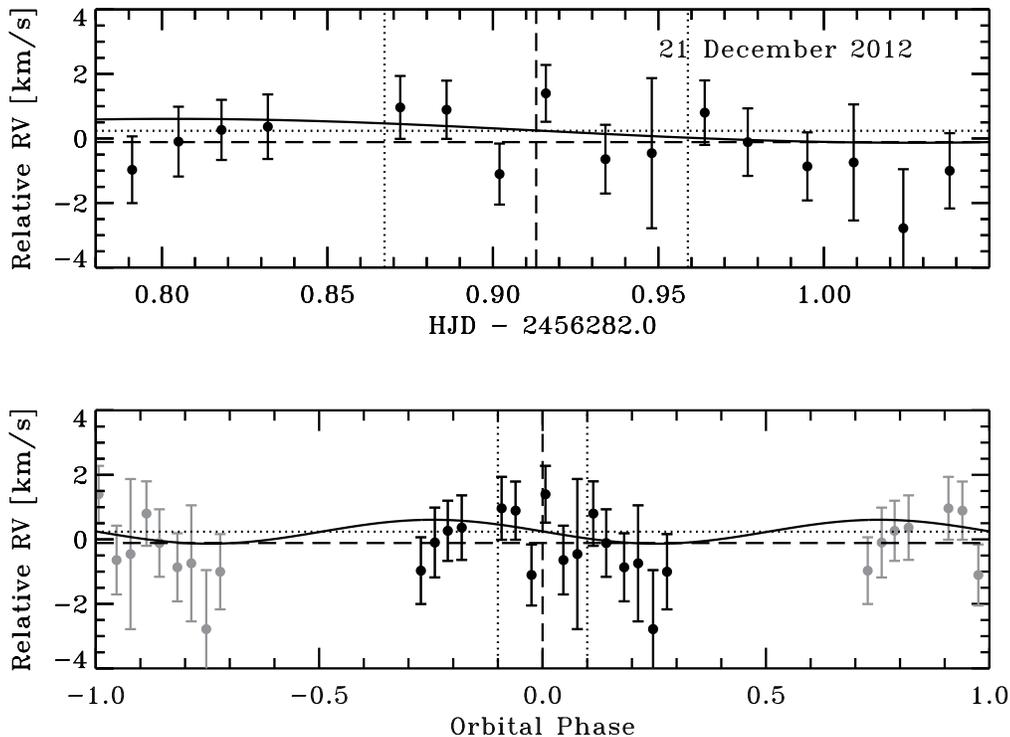}

    \figcaption{\emph{Top:} December 2012 Keck NIRSPEC radial velocity data 
plotted as a function of time.  The horizontal dashed line represents the 
weighted mean of the relative velocities and the solid black line represents the 
best fit radial velocity curve with the transit midpoint (vertical dashed line) 
fixed by the Spitzer ephemeris (Section~\ref{ephm-sub}); the horizontal dotted 
line represents the velocity offset of the fit ($\gamma = 0.237\pm0.415$ km/s). 
The vertical dotted lines represent the predicted beginning and ending times 
of ingress and egress, respectively. \emph{Bottom:} The radial velocity curve 
is phased on the Spitzer ephemeris.  The overplotted lines represent the fits as 
in the top panel; the gray points are the values repeated in phase for 
continuity and clarity. \label{keckrv-fig}}

\end{figure*}
\section{Observations and Analysis}\label{observation-sec}
\begin{figure*}[ht]
     \centering
    \includegraphics[angle=90,scale=0.6,keepaspectratio=true]{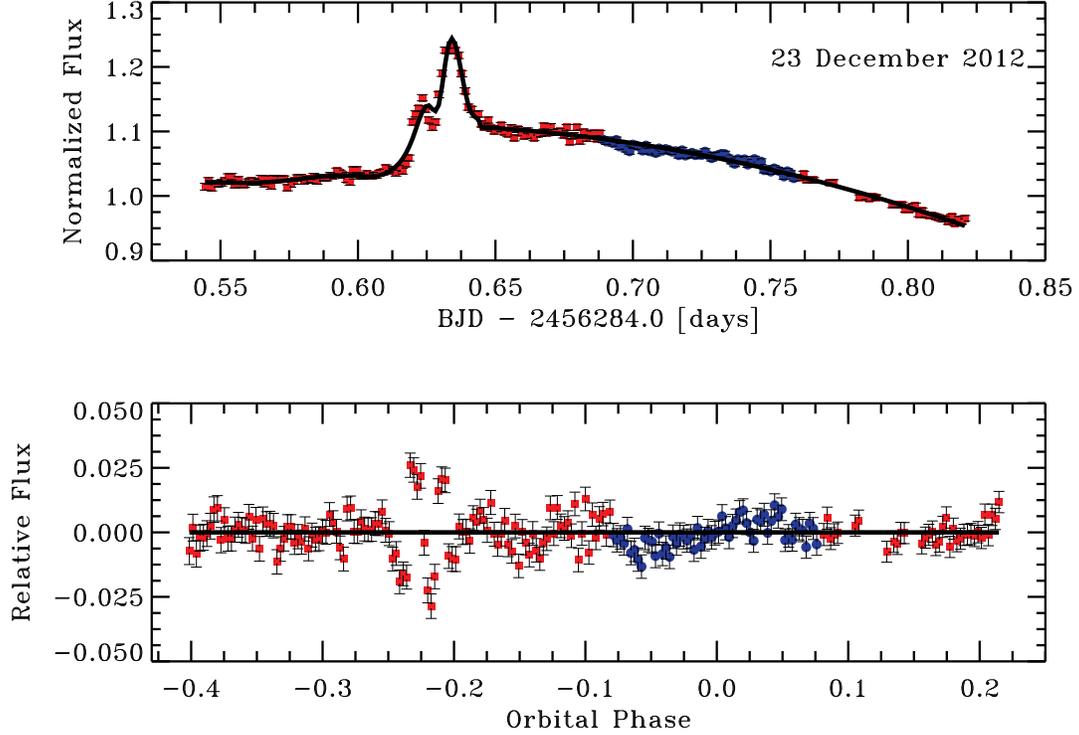}

    \figcaption{\emph{Top:} December 2012 LCOGT $r^\prime$ light curve, obtained 
two days after the Keck NIRSPEC observations, is plotted as a function of time. 
The blue circles indicate the data points predicted to be within the transit 
window from the Spitzer ephemeris (Table~\ref{ephm-tab}); the red squares are 
outside the transit window and were used to fit a cubic spline (black line) to 
model the stellar variability. \emph{Bottom:} The LCOGT $r^\prime$ light curve, 
phased upon the Spitzer ephemeris and with the spline fit subtracted, shows no 
sign of a transit. The solid black line represents a constant value of 0.0. 
\label{lconontransit-fig}}

\end{figure*}
\begin{figure*}[ht]
     \centering
    \includegraphics[angle=90,scale=0.6,keepaspectratio=true]{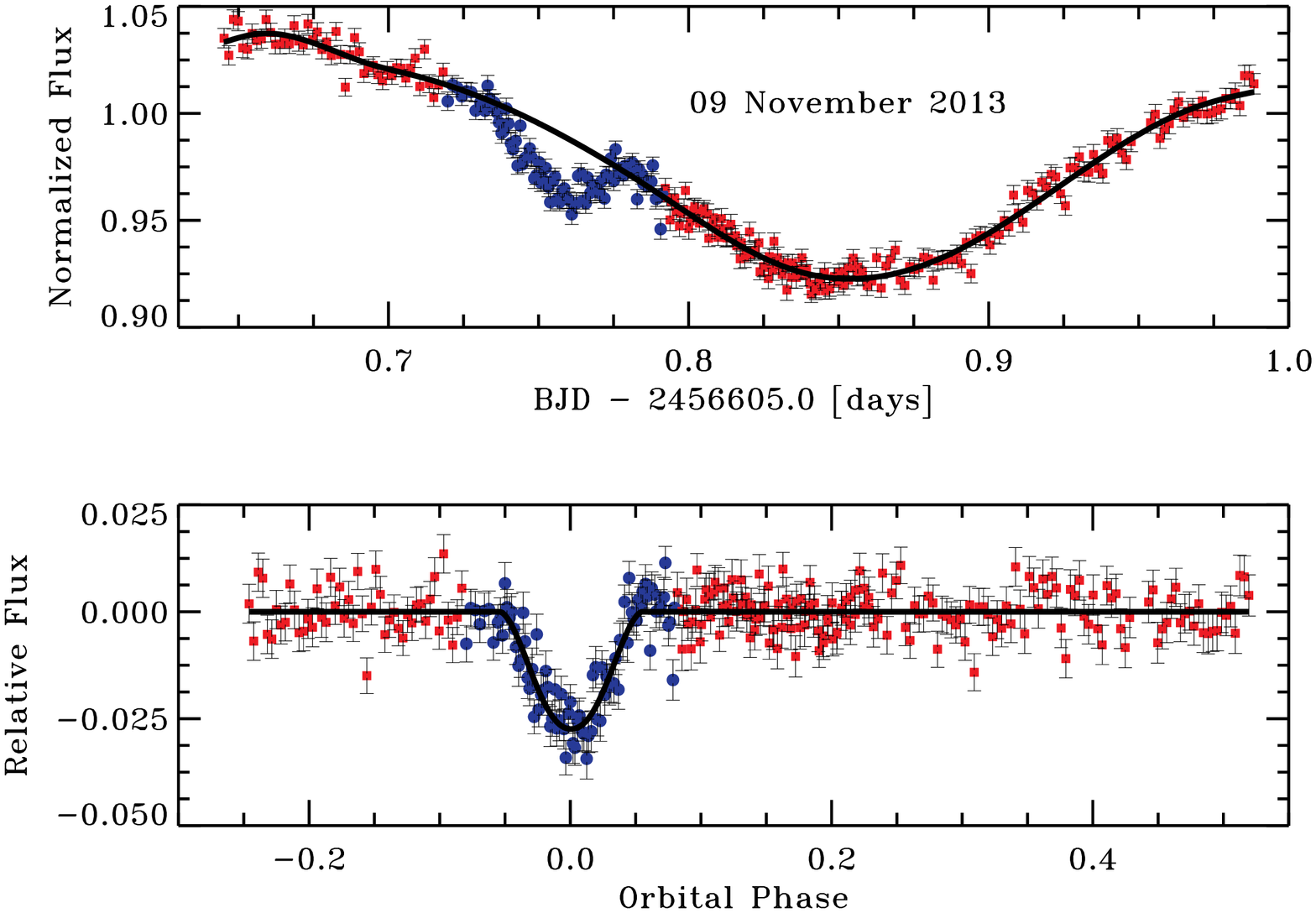}
     \figcaption{\emph{Top:} November 2013 LCOGT $r^\prime$ light curve, 
obtained approximately 3 years after the PTF data and 1.5 years after the 
Spitzer data, is plotted as a function of time. The blue circles indicate the 
data predicted to be within the transit window from the Spitzer ephemeris 
(Table~\ref{ephm-tab}); the red points are outside the transit window and were 
used to fit a $8^{th}$-order polynomial (black line) to model the stellar 
variability. \emph{Bottom:} The LCOGT light curve, with the polynomial fit 
subtracted to remove the stellar variability, is plotted as a function of 
orbital phase. The solid black line represents the transit fit to the light 
curve and the data have been phased on this new ephemeris (see 
Table~\ref{ephm-tab}).\label{lcotransit-fig}}

\end{figure*}
\subsection{Spitzer 4.5\micron}\label{observation-spitzer-sub}

Spitzer Space Telescope Director's discretionary time was granted to observe 
PTFO~8-8695 on 28 April 2012 UT, soon after the discovery paper was published. 
The primary purpose of the observations was to confirm the presence of the 
transit two years after the PTF Orion data were acquired \citep{vaneyken12}, and 
to search for a secondary eclipse. The infrared observations of PTFO~8-8695 
consist of an 11.8 hour stare at 4.5\micron\ with IRAC (request key 45476608), 
consisting of 1368 thirty second integrations with an effective cadence of 
$\sim$ 31.3 seconds.   The data were collected using the standard techniques for 
high-precision photometry with IRAC to minimize instrumental systematics 
\citep{ingalls12}.  The observations were placed on the part of an array pixel 
that has minimal response variations.  The standard Spitzer Science Center data 
products were used in our analysis.

Aperture photometry was performed on each basic calibrated data (BCD) image 
using a three pixel radius aperture and subtracting a background using an 
estimate of the mode of a circular annulus with inner radius of 3 and outer 
radius of 7.  Source positions were determined on each BCD by calculating the 
center-of-light using a $5\times5$ pixel box centered on the peak pixel 
associated with the source using the Spitzer Science Center provided IDL 
procedure box\_centroid.pro; aperture photometry was performed utilizing the IDL 
version of the DAOPHOT routine aper.pro. Photometric systematics were removed by 
applying an intra-pixel gain map \citep{ingalls12} to each photometric point as 
a function of centroid position. The Spitzer light curve is tabulated in 
Table~\ref{spitzerlc-tab} and shown in Figure~\ref{spitzer-fig}.
\begin{deluxetable}{crrc}
\tablecolumns{4}
\tablecaption{Keck NIRSPEC Radial Velocity Data\label{keckrv-tab}}
\tablehead{
\colhead{HJD} &
\colhead{Phase\tablenotemark{a}} &
\colhead{Radial Velocity} & 
\colhead{Uncertainty}\\
\colhead{}&
\colhead{}&
\colhead{[km/s]}&
\colhead{[km/s]}
}
\startdata
2456282.791  &  $-0.272$ & $-0.971$ & 1.034\\
2456282.805  &  $-0.241$ & $-0.098$ & 1.082\\
2456282.818  &  $-0.212$ & $0.265$  & 0.932\\
2456282.832  &  $-0.181$ & $0.364$  & 1.000\\
2456282.872  &  $-0.092$ & $0.963$  & 0.975\\
2456282.886  &  $-0.060$ & $0.890$  & 0.902\\
2456282.902  &  $-0.024$ & $-1.104$ & 0.945\\
2456282.916  &  $0.007$  & $1.398$  & 0.881\\
2456282.934  &  $0.047$  & $-0.642$ & 1.065\\
2456282.948  &  $0.078$  & $-0.456$ & 2.327\\
2456282.964  &  $0.114$  & $ 0.800$ & 0.999\\
2456282.977  &  $0.143$  & $-0.114$ & 1.046\\
2456282.995  &  $0.183$  & $-0.866$ & 1.055\\
2456283.009  &  $0.214$  & $-0.743$ & 1.798\\
2456283.024  &  $0.247$  & $-2.783$ & 1.828\\
2456283.038  &  $0.279$  & $-1.003$ & 1.167
\enddata
\tablenotetext{a}{Based upon Spitzer Ephemeris (Table~\ref{ephm-tab}).}
\end{deluxetable}
\subsection{Keck-II NIRSPEC Radial Velocities}\label{observation-keck-sub}

Keck-II NIRSPEC observations were obtained on 21 December 2012 -- a full two 
years after the detection of the optical transits with PTF.  The purpose of the 
NIRSPEC observations was to obtain phase-resolved high precision radial velocity 
data,  primarily during the transit of PTFO~8-8685b.  The Rossiter-McLaughlin 
effect was predicted to be $\approx 2-3$ km/s given the $v\sin i$ of the star 
and the depth of the transit \citep{vaneyken12}.

Spectra were acquired through the $0.432^{\prime\prime}$ (3 pixel) wide slit; 
the echelle and cross disperser angles were oriented to obtain K-band spectra 
containing Na I, MgI and CO. In this configuration, orders 32-38 
(non-contiguous) were imaged through the NIRSPEC-7 blocking filter.  The data 
were acquired in pairs of exposures of 600 seconds in a standard ABBA sequence, 
with the telescope being nodded $\pm6^{\prime\prime}$ from the slit center along 
the $24^{\prime\prime}$ slit, so that each frame pair would contain object and 
sky in both nod positions. 

To correct for telluric absorption, we also obtained spectra of HR 2315, an A0V 
star located close to PTFO~8-8695 in the sky, before and after the PTFO~8-8685 
observations. After every two ABBA sets, spectra of the internal NIRSPEC 
continuum lamp were taken for flat fields at the K-band settings; lamp 
exposures of the argon, neon, krypton, and xenon arc lamps provided wavelength 
calibration for all the K-band orders. At the end of the PTFO~8-8695 science 
observations, GJ281, a radial velocity standard, was observed to set the 
absolute velocity scale of the observations.   The spectral images were dark 
subtracted and flat fielded and the individual nods were extracted, wavelength 
calibrated with the lamp spectra, and telluric divided.  Four ABBA nod 
sets were obtained for a total of 16 spectra and radial velocity estimates. The 
night was plagued with variable high cirrus and as a result,  the 
signal-to-noise per resolution element varied but was approximately 50.

The Na I lines at $\sim2.206\micron$ are the strongest lines in the 
near-infrared spectrum of PTFO~8-8695 and the entire order containing these 
lines was used to determine the relative radial velocities. The telluric 
corrected individual spectra were coadded to produce a master spectrum with 
$S/N \sim 200$.  The master spectrum was cross correlated with each of the 16 
individual spectra to determine the relative radial velocities between the 
spectra. A Monte Carlo effort was employed to estimate the uncertainties.  The 
Monte Carlo simulations assume that the wavelength fits are perfect, which, of 
course, may not be correct.  To try to characterize the wavelength fitting 
uncertainty, the spectra with strong telluric lines were utilized. The rms 
scatter in the radial velocity measurements of the telluric lines in each of 
the 16 spectra was found to be 0.68 km/s. This uncertainty contribution was 
added in quadrature to the Monte Carlo uncertainty to arrive at a final 
uncertainty.  The measured radial velocities and associated uncertainties are 
presented in Table~\ref{keckrv-tab}, and the radial velocity curve is shown in 
Figure~\ref{keckrv-fig}.

\begin{deluxetable*}{ccccccc}[b]
\tablecolumns{7}
\tablewidth{6in}
\tablecaption{LCOGT Light Curve Data (sample)\label{lcogtlc-tab}}
\tablehead{
\colhead{BJD} &
\colhead{Phase} & 
\colhead{Flux} &
\colhead{Uncertainty} &
\colhead{Flux Model} &
\colhead{Flux$-$Model} &
\colhead{Transit Model}
}
\startdata
2012 December \\
2456284.54447 & -0.4014 &  1.0143089 & 0.0052217 &  1.0215716 & -0.0072627 & 
\nodata \\
2456284.54568 & -0.3987 &  1.0232391 & 0.0051967 &  1.0214167 &  0.0018223 & 
\nodata \\
2456284.54687 & -0.3960 &  1.0129543 & 0.0051269 &  1.0212653 & -0.0083110 & 
\nodata \\
2456284.54807 & -0.3933 &  1.0188872 & 0.0050400 &  1.0211184 & -0.0022312 & 
\nodata \\
& \\
2013 November \\
2456605.75391 & -0.0035 &  0.9584972 & 0.0040311 &  0.9926078 & 
-0.0341105 & 
-0.0272040 \\
2456605.75483 & -0.0014 &  0.9682513 & 0.0047093 &  0.9919295 & -0.0236782 & 
-0.0273680 \\
2456605.75546 &  0.0000 &  0.9703735 & 0.0040732 &  0.9914676 & -0.0210941 & 
-0.0274150 \\
2456605.75639 &  0.0021 &  0.9600239 & 0.0046811 &  0.9907754 & -0.0307514 & 
-0.0273890
\enddata
\end{deluxetable*}
\subsection{LCOGT Photometry}\label{observation-lco-sub}
We obtained optical photometry for PTFO~8-8695 using the LCOGT 1m telescope 
network during December 2012 and November 2013. We utilized the full network 
capabilities of LCOGT including the sites at Cerro Tololo Inter-American 
Observatory in Chile (CTIO), Siding Spring Observatory in Australia (SSO), the 
South African Astronomical Observatory at Sutherland, South Africa (SAAO), and 
McDonald Observatory in Texas (MDO). The December 2012 data were obtained in 
support of the Keck NIRSPEC observations and were intended to be simultaneous 
with the Keck Observations, but poor weather prevented data from being obtained 
on the exact night of observations. Instead, contemporaneous  data were obtained 
on 23 December 2012 UT. Exposure times were 90 sec, resulting in a cadence of 
approximately 103 seconds. All exposures were taken in the $r^\prime$ filter, 
similar to the filter used in the PTF--Orion survey data.  

After \citet{barnes13} published the precession model, it was recognized that 
the orbital plane could precess causing the transits to change, disappear and 
reappear, LCOGT was utilized again on 09 November 2013 UT to observe PTFO~8-8695 
data in order to confirm the reappearance of the transit events and, in 
conjunction with the Spitzer photometry and the Keck radial velocity data, 
confirm the precession model and the planetary nature of PTFO~8-8695b. The 
November 2013 data were also obtained in the $r^\prime$ filter with an exposure 
time of 120 seconds; because the network of telescopes was able to observe the 
target with multiple telescopes at the same time, the effective cadence of the 
observations ranged from $\approx$2 seconds to $\approx$130 seconds with the 
majority ($\approx$ 50\%) of the observations obtained at a 130 second cadence.

In order to create the differential photometry, an initial normalization curve 
was created by taking the simple mean in magnitude space of all the raw 
reference light curves. For each reference star light curve, the mode of the 
residuals against the normalization curve was then subtracted, so that all the 
reference star light curves were normalized to the same flux level. Exposure by 
exposure, the mode of all the now-normalized reference star magnitudes was 
found, yielding the differential offset correction needed for each exposure. 
These differential corrections were subtracted from the raw target light curve 
to produce the final differentially-corrected photometry. The same corrections 
applied to the original reference stars themselves (which should yield flat 
light curves) provided an internal consistency check. For a more detailed 
overview of the differential photometry technique, see \citet{vaneyken11}.  The 
LCOGT light curves are tabulated in Table~\ref{lcogtlc-tab} and shown in 
Figures~\ref{lconontransit-fig} and \ref{lcotransit-fig}.

\section{Discussion}\label{discussion-sec}

\subsection{Transit Ephemerides}\label{ephm-sub}
\begin{deluxetable*}{lccc}
\tablecolumns{4}
\tablewidth{6.0in}
\tablecaption{Transit Ephemerides\label{ephm-tab}}
\tablehead{
\colhead{} &
\colhead{PTF\tablenotemark{a}} & 
\colhead{Spitzer} & 
\colhead{LCOGT}
}
\startdata
Approximate Date of Data Collection& 2009 Dec \& 2010 Dec & 2012 April & 2013 
November\\
Transit Midpoint [Reduced BJD]\tablenotemark{b} & 5543.9402 & 6045.7026 & 
6605.7555\\
Midpoint Uncertainty & $\pm 0.0008$ & $\pm 0.0009$ & $\pm 0.0004$\\
Transit Depth & $3-5$\% & $0.58\pm0.01$\% & $2.74\pm0.01$\%\\
Days Past PTF Midpoint & \nodata & 501.7622 & 1061.8149\\
Number of Orbits since PTF Midpoint\tablenotemark{c} & \nodata & 1119 & 2368\\
Uncertainty in Transit Prediction from PTF Midpoint& \nodata & $\pm 65$ min. & 
$\pm 136$ min.\\
Measured Offset from PTF Midpoint Prediction & \nodata & $-17.2$ min. & $-38.9$ 
min.\\
Days Past Spitzer Midpoint & \nodata & \nodata & 560.0527\\
Number of Orbits since Spitzer Midpoint\tablenotemark{c} & \nodata & \nodata & 
1249\\
Uncertainty in Transit Prediction from Spitzer Midpoint & \nodata & \nodata & 
$\pm 71$ min.\\
Measured Offset from Spitzer Midpoint Prediction & \nodata & \nodata & $-21.8$ 
min.

\enddata
\tablenotetext{a}{From \citet{vaneyken12,barnes13}}
\tablenotetext{b}{Reduced BJD = BJD - 2450000}
\tablenotetext{c}{Period held constant at $P = 0.448413\pm0.000040$ days.}

\end{deluxetable*}

The Spitzer light curve was acquired $\sim 502$ days past the time of the 2010 
PTF transit midpoint determination; in that time, the planet candidate 
PTFO~8-8695b would have orbited its host star  1119 times.  As a result, the 
timing of the transit in the Spitzer data was uncertain by 65 minutes.  The 
transit was identified by visual inspection of the light curve at the time 
predicted by the PTF ephemeris. The transit event was found to be within the 
uncertainties of the predictions, but to locate the transit more precisely, we 
excluded those data within $\sim$2.5 hours of the predicted transit time 
(Fig.~\ref{spitzer-fig} blue circles) and fit an $8^{th}$-order polynomial to 
the data outside the transit window (Fig.~\ref{spitzer-fig} red squares). Given 
the large number of data points in the Spitzer light curve outside the transit 
window (1174) and the fact that the polynomial is only intended to 
parametrically model the stellar variability, the order of the polynomial was 
set by requiring a non-significant change in the normalized variance of the 
residuals.  The polynomial fit was subtracted from the Spitzer light curve (data 
in the transit window included) to produce a whitened light curve suitable for a 
transit model fit (bottom of Fig.\ref{spitzer-fig}).  

The Spitzer light curve was fit using EXOFAST \citep{eastman13} which yielded a 
value and uncertainties via MCMC modeling for the Spitzer transit midpoint and 
the transit depth (see Table~\ref{ephm-tab}).  The Spitzer transit was found to 
be 17 minutes earlier than predicted, well within the 65 minute uncertainty from 
the 2010 PTF ephemeris uncertainty propagated to the time of the 2012 Spitzer 
observations.  Thus, more than 16 months after the PTF observations,  the 
transit event occurred close to the predicted time.   However, the transit, 
unlike the optical PTF transits, was found to be only $\sim 0.6$\% deep.

One of the primary purposes of the Spitzer observations was to search for a 
secondary eclipse; however, we found no evidence for a secondary eclipse to 
within the limits of the data.  At a semi-major axis distance of 0.00838 AU from 
the star ($T_{eff} = 3470$K, $R_\star = 1.4\ R_\odot$), the planet candidate 
should have an equilibrium temperature of $T_{eq} \approx 800 - 1800$K, 
depending upon the albedo of the planet - similar to the expected effective 
temperature of a few million year old Jupiter-mass planet \citep{baraffe03}.  
The Spitzer 4.5\micron\ light curve places a limit on the depth of the secondary 
eclipse of $\lesssim 0.3$\% ($3\sigma$).  This translates to an upper limit on 
the planetary candidate radius of $R_p \lesssim 1.5\ R_{Jup}$, in reasonable 
agreement with the radius derived directly from the transit depths.

The 2012 December LCOGT optical light curve was obtained 8 months after the 
Spitzer light curve, in support of the Keck radial velocity data.  In that time, 
the candidate planet PTFO~8-8695b would have orbited an additional 530 times and 
have a predicted transit time uncertainty of $\sim 10$ minutes (0.02 in phase). 
As with the Spitzer light curve, data within $\sim$2.5 hours of the predicted 
transit midpoint were excluded, and the out-of-transit data points were fit with 
a cubic spline;  the flux brightening event that resembles a flare  at 
BJD=2456284.625 prevented the use of a polynomial to parameterize the 
out-of-transit variability (Fig.~\ref{lconontransit-fig}).   After subtraction 
of the spline fit, the light curve shows no clear sign of a transit event to a 
limit of $\sim 0.7\%\  (1\sigma$) within 0.02 in phase of the predicted transit 
time.  Structure is seen throughout the residual light curve at all phases and 
is likely the result of correlated noise in the data and/or true variability in 
the star.

The 2013 November LCOGT optical light curve was obtained 560 days after the 
Spitzer light curve, as part of an effort to re-detect the transits -- a 
necessary link  in our efforts to confirm the orbital precession 
model.   In that time, the  the planetary candidate 
would have orbited an additional 1249 times and have  a predicted 
transit time uncertainty of about 71 minutes. Even without the formal searching, 
the transit event is clearly seen at the time of predicted transit (see 
Fig.~\ref{lcotransit-fig}). To be consistent with our analysis of the Spitzer 
light curve and the previous LCOGT light curve, data within $\sim$2.5 hours of 
the predicted transit window were excluded (Fig.~\ref{lcotransit-fig} blue 
circles) and an $8^{th}$-order polynomial was fit to the data outside the 
transit window (Fig.~\ref{lcotransit-fig} red squares). The polynomial fit was 
subtracted from the LCOGT light curve (including data within the transit window) 
to produce a whitened light curve suitable for a transit model fit (bottom of 
Fig.\ref{lcotransit-fig}). 

The light curve was fit using EXOFAST which yielded a value for the LCOGT 
transit midpoint and the transit depth (see Table~\ref{ephm-tab}).  The transit 
was found 21 minutes earlier than predicted but again well within the 71 minute 
uncertainty from the Spitzer ephemeris uncertainty propagated to the time of the 
LCOGT observations; in fact, the measured time of transit was only 39 minutes 
earlier than what was predicted from the PTF ephemeris which was established 3 
years and 2368 orbits earlier. Thus, the planetary candidate PTFO~8-8685b 
transited the star as predicted  via simple extension of the transit 
timing ephemerides; this transit has a measured depth of $\sim 
2.7$\%.

As a final note on the stability of the transit ephemeris, the period was 
originally determined to within 3.5 seconds \citep{vaneyken12}.  If the measured 
transit midpoints for the PTF, Spitzer, and LCOGT light curves were perfect, the 
period uncertainty, propagated to the dates of the observations, would still 
produce an uncertainty greater than the measured offsets of the observed Spitzer 
and LCOGT transits. If the period was shorter by only 0.98 seconds, the transits 
would have been predicted at the times observed, indicating the quality of the 
original  measured  ephemerides. 

\subsection{Radial Velocity Limits}\label{rvfit-sub}
When originally proposed, the 2012 December Keck NIRSPEC radial velocity 
observations were intended to measure the Rossiter-McLaughlin effect as the 
planet transited the star.  The observations were timed to be centered on the 
transit midpoint, as predicted from the PTF ephemeris.  The amplitude and shape 
of the Rossiter-McLaughlin effect depends heavily on the $v\sin i$ of the star, 
the size of the transiting planet, the orbital geometry of the system, and the 
viewing geometry from the Earth. We predicted that given the transit depth of 
$\sim 4$\% and a stellar $v\sin i \approx 80$ km/s, the amplitude of the 
Rossiter-McLaughlin effect should be $2-4$ km/s above the nominal radial 
velocity curve.  

From Figure~\ref{keckrv-fig}, it is clear that the Rossiter-McLaughlin effect 
was not detected, nor was the radial velocity signature of the  planetary 
candidate PTFO~8-8695b  detected within the limits of the Keck NIRSPEC data. The 
radial velocity curve is consistent with a constant value with a reduced 
chi-squared $\chi^2 = 0.85$. The lack of detection of the Rossiter-McLaughlin 
effect is consistent with the LCOGT light curve from 2012 December  which showed 
that there was not a transiting event at the time of the Keck observations. 

 We have fitted the radial velocities with a Keplerian orbit, where 
the orbital period was fixed to 0.448413 days from \citet{vaneyken12}, and the 
orbital eccentricity was set to 0.0 consistent with both \citet{vaneyken12} and 
\citet{barnes13}.  The time of the inferior conjunction (i.e., 
transit) was fixed by the Spitzer ephemeris listed in Table~\ref{ephm-tab}.  The 
Keplerian orbital solution to the radial velocity data in Table~\ref{keckrv-tab} 
used RVLIN, a partially linearized, least-squares fitting procedure described in 
\citet{wh09}; parameter uncertainties were estimated using the BOOTTRAN 
bootstrapping routines described in \citet{wang12}. The results of the fitting 
are plotted over top the observations in Fig.~\ref{keckrv-fig}.  The Keplerian 
orbital fit does not improve significantly the fit over the constant weighted 
mean with only a reduced chi-squared of $\chi^2 = 0.80$, and the semi-amplitude 
of the fit is statistically consistent with a flat line ($K = 0.37 \pm 0.33$ 
km/s).    

\begin{figure*}[ht]
     \centering
    \includegraphics[angle=0,scale=0.45,keepaspectratio=true]{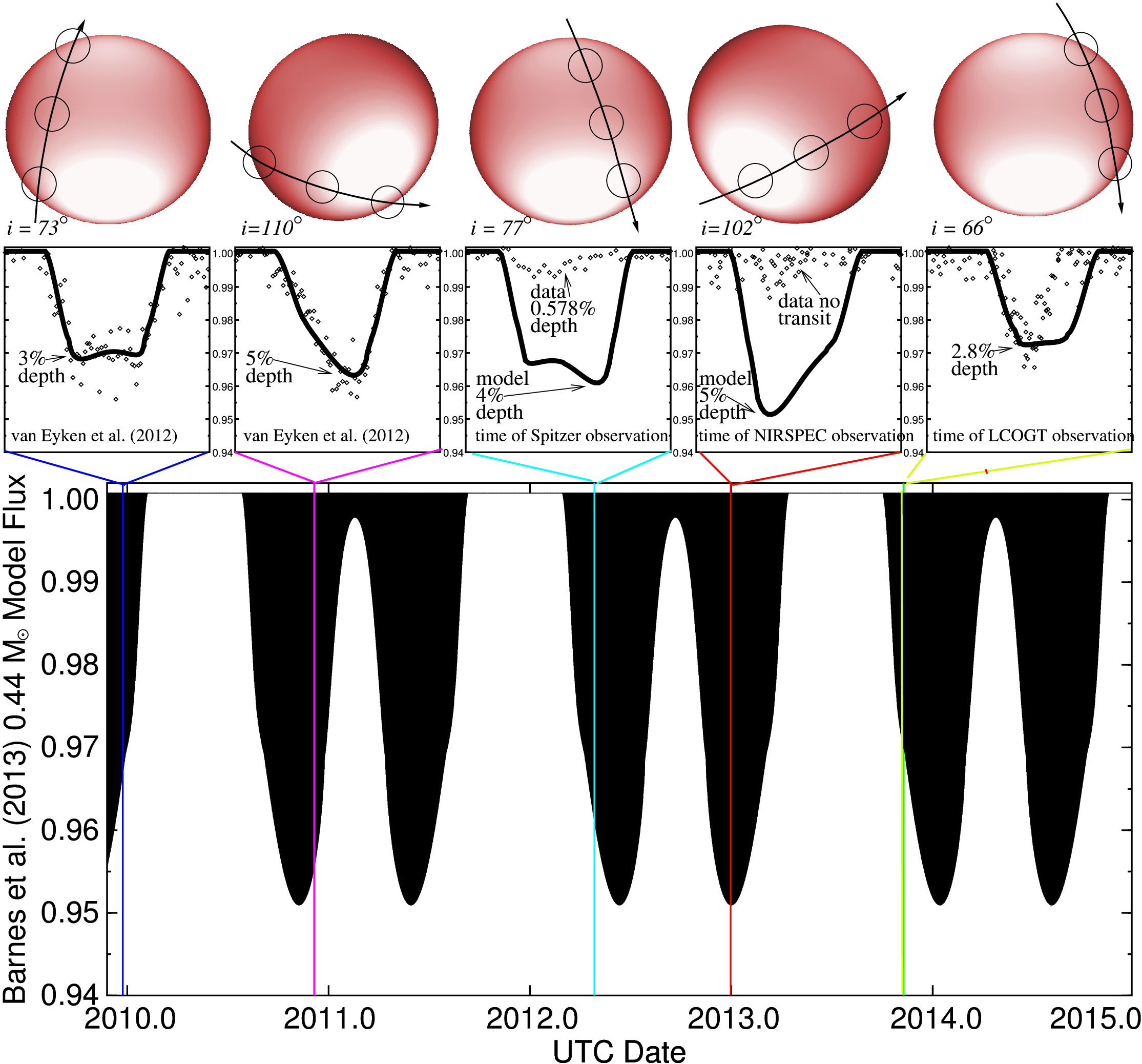}

    \figcaption{The precession model from \citet{barnes13} for a stellar mass of 
$M_\star = 0.44\ M_\odot$ is compared to the observations.  The bottom panel 
shows the predicted transit depths as a function of time starting with the 2009 
PTF season and projected out through the end of 2014.  The middle panel shows 
the transit model predictions at each of the times of observation.  The PTF 2009 
and PTF 2010 transits match the model by default as these transits were used as 
constraints in the models.  The transit model predictions at the time of the 
Spitzer, Keck/LCOGT, and LCOGT observations are shown.  The top panel shows the 
orientation of the star and the planet during the time of observation. The 
shading on the star represents the brightness gradient caused by the gravity 
darkening; white is brighter and represents the poles of the star. The 
line-of-sight inclination ($i$) at the time of the observation is also listed.  
Inclinations greater than $90^\circ$ are related to the orientation of the 
system as a whole \citep[for a description of the geometry see][]{barnes13}. 
\label{precession0.44-fig}}
%
     \centering
    \includegraphics[angle=0,scale=0.45,keepaspectratio=true]{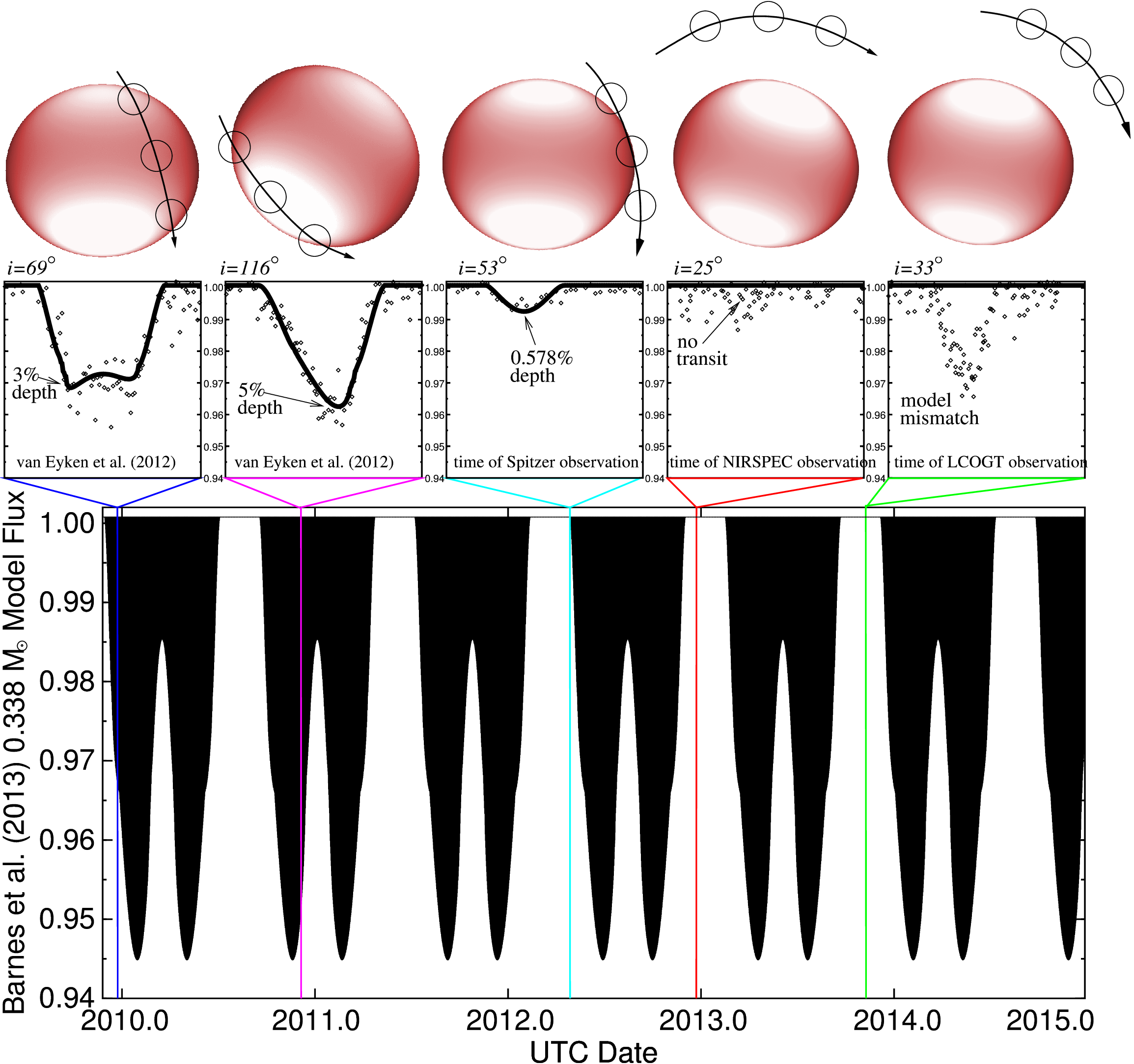}
     \figcaption{The precession model from \citet{barnes13} for a stellar mass 
of $M_\star = 0.34\ M_\odot$ is compared to the observations.  The figure 
content is the same as for Figure~\ref{precession0.44-fig}.   
\label{precession0.34-fig}}

\end{figure*}

\subsection{Comparison to the Precession Models}\label{models-sub}

The mass of the host star PTFO~8-8695 was estimated originally from isochrone 
fitting \citep{briceno05}, but the data could not distinguish between models 
from \citet{baraffe10} and \citet{siess00}; thus, leaving an ambiguity in the 
stellar mass of either $\sim 0.44\ M_\odot$ \citep{baraffe10} or $\sim 0.34\ 
M_\odot$ \citep{siess00}.  As a result, \citet{barnes13} developed a set of 
models for both stellar mass estimates and fit the observed PTF transits from 
2009 and 2010 separately and simultaneously.  The separate models described the 
expected rotational axis and orbital planet precession of the star and planet, 
and predicted the depth, shape, and times of the planetary transits for each of 
the stellar masses. Additionally, the precession modeling does take into account 
the wavelength dependence of the transits - in particular, the effects of 
weakened limb and gravity darkening at the wavelengths of the Spitzer 
observations \citep{barnes13}. The data presented here are compared to the two 
models from \citet{barnes13}.

The higher stellar mass ($0.44\ M_\odot$) precession model predicts a precession 
period of 581 days with a 184 day stretch with no transits visible at all as a 
result of the orbit precessing out of the line of sight.  Within the time that 
transits are visible, there is a double peak of deep transits separated by about 
200 days.  The lower mass ($0.34\ M_\odot$) precession model predicts a similar 
transit structure as the higher mass model but on a more condensed time scale 
(Figure~\ref{precession0.34-fig}). The precession period is approximately 293 
days and the time between the deep transits is about 80 days and the transits 
only disappear completely for approximately 75 days.

Overall, the observations do not agree well with the $0.44\ M_\odot$ model. The 
Spitzer observations, observed to be only $\sim 0.6$\% in depth are predicted by 
the model to be $\sim 4$\%.  Further, the observations in 2012 December are 
predicted to have occurred when the transit depth was at its deepest $\sim 5$\%. 
Such a deep $5$\% deep transit would have been detected in the LCOGT data which 
have a transit detection limit of $\sim 0.7$\% (Fig.~\ref{lconontransit-fig}).  
Additionally, with a line-of-sight inclination of $i=102^\circ$, the amplitude 
of the Rossiter-McLaughlin effect should have been $\gtrsim2.5$ km/s and would 
likely have been detected by the NIRSPEC radial velocity data at $>2\sigma$.  
Finally, the 2013 November optical transit detection is predicted at 
approximately the right depth, but the shape of the observed transit is much 
narrower in comparison to the predicted transit, indicating that the model 
incorrectly represents the system orientation at the time of the observations. 
Thus, the $0.44\ M_\odot$ precession model is inconsistent with the observations 
for each of the three epochs of data presented.

The $0.34\ M_\odot$ precession model somewhat more closely matches the 
observations (Fig.~\ref{lcotransit-fig}).  The Spitzer light curve is 
represented well by the precession model transit prediction. Both the depth and 
the shape of the observed transit are in good agreement with the model 
prediction. Further, at the epoch of observations for the Keck NIRSPEC data and 
the first LCOGT light curve (2012 December), the model predicts that the system 
should be non-transiting, and, indeed, the observations support the 
disappearance of the transiting events.  The model predicts the return of the 
transits in late 2013, and the 2013 November data confirm that the transits do 
reappear. However, the observations showed that the transits reappeared about 1 
month earlier than predicted. With a precession orbital period of 293 days, the 
model is out-of-sync with the observations by $\sim10-15$\% of the predicted 
precession period.

The cause of the inconsistency of the $0.34\ M_\odot$ model with the 2013 
November observations, while predicting the 2012 April and 2012 December 
observations correctly, is difficult to discern, but could be the result of a 
variety of factors.  The precession model depends heavily on knowledge of the 
stellar mass,  the stellar moment of inertia, the stellar radius, and the 
planetary orbit eccentricity.  The precession model assumes a circular orbit 
with a stellar radius of $\sim 1.04\ R_\odot$; however, an eccentric orbit would 
allow for a larger stellar radius, which would affect the details of the 
precession model and the precession period. The precession period is, to first 
order, proportional to the square of the stellar radius ($R_\star^2$), and is 
also dependent on higher order powers of the stellar radius (e.g., $R_\star^4$) 
\citep{md00, barnes13}; a $5-7$\% change in the stellar radius could result in a 
$10-15$\% change in the precession period.

The precession model also uses as inputs the orbital period and epochs of 
transits; even small errors on these, when propagated across three years and 
thousands of orbits, may explain why the $0.34\ M_\odot$ model predicts 
correctly the 2012 data but not the 2013 data.  Finally, the precession modeling 
fits not only the timing of the transits but also the shape of the transits.  If 
spots have significantly altered the shape of the transit, the precession model 
may not correctly predict the system parameters -- particularly when the model 
is propagated through thousands of orbits. 

We cannot place better constraints on the stellar or planetary candidate 
parameters with the data presented here.  The parameter space of the precession 
models is extremely large and is difficult to narrow and beyond the intended 
scope of this paper. A recent paper, studying the same system with our PTF data 
but without requiring the spin-orbit to be synchronously locked, find that a 
precession model can still reproduce the observations, with a precession period 
of 199 days and a planet mass of $\sim 4-5 M_{Jup}$ \citep{kamiaka15}.   We are 
continuing to observe the planetary candidate with long-term photometric 
observations and radial velocity monitoring in an effort to assess more fully 
the existence and planetary nature of PTFO~8-8695b and to more carefully 
determine the validity and parameters of the precession model.

While the observations presented here cannot uniquely identify which precession 
model might best describe the data, we can utilize the NIRSPEC radial velocity 
upper limits on the semi-amplitude to set limits on the potential mass of the 
planetary candidate. If the stellar mass is $0.44\ M_\odot$, then the mass limit 
on a possible planetary candidate is $M_p\sin i \lesssim 0.8\ M_{Jup}$; if the 
stellar mass is closer to $0.34\ M_\odot$, then the mass limit is $M_p\sin i 
\lesssim 0.7\ M_{Jup}$.  As indicated by the data, the planet candidate does not 
transit during the time of the radial velocity observations, which sets an upper 
limit on the line-of-sight orbital inclination of $i \lesssim 50^\circ$. At that 
inclination, the planet candidate mass would have a mass of $M_p \lesssim 0.9 - 
1.1\ M_{Jup}$. If the orbital inclination is as low as $i \sim 
20^\circ-30^\circ$ as predicted by the $0.34\ M_\odot$ precession model, then 
the planetary candidate could have a mass of $M_p \lesssim 2.0 - 2.5\ M_{Jup}$.

\subsection{Stellar Activity}
PTFO~8-8685 is a low mass young star and as a result is variable and is likely 
spotted as discussed in \citet{vaneyken12}\citep[see also][]{koen15}. The 
optical radial velocity curve modulations were attributed to the presence of 
spots, although not necessarily to the presence of a single spot.   Large and 
long-lived, high latitude spots are commonly present in weak-lined T Tauri stars 
\citep[e.g.,][]{stelzer03,rice11}, and spots have been seen to mimic eclipsing 
companions in other systems \citep[e.g., RXJ1608.6-3922;][]{joergens01}.  In 
that work, light curve data obtained in multiple filters and spectroscopic 
monitoring enabled the authors to determine that the deep eclipses (0.5 mag), 
with long eclipse durations ($\sim 0.5$ in phase), were caused by stellar spots 
which disappeared in observations obtained 4 years later.  They also found that 
the photometric variations were also consistent with the 2.4 km/s radial 
velocity variations but not compatible with an eclipsing binary. 
  
\cite{vaneyken12} could not find a spot distribution solution to fit the 
observed light curves for PTFO~8-8695, but we revisit the possibility of spots 
as a cause of the observed transit events here.  Unlike RXJ1608, the observed 
transit features for PTFO~8-8695 are more shallow ($3\% - 5\%$) and shorter in 
duration (0.1 in phase), and, as seen in 2009, sometimes flat-bottom during the 
event.  In principle, a spotted surface with a single hot spot near the pole 
briefly eclipsed by the limb of the star may be able produce the short transit 
duration observed.  However, such a hot spot would be expected to show a 
brightening $180^\circ$ out of phase with the flux dip but such a brightening is 
not seen in the data.  In fact, many of the light curves from the 2009/2010 
discovery data \citep{vaneyken12} display a brightness decrease near these 
phases in the light curves.

A cold spot below the equator could briefly come into view. In general, the spot 
size (i.e., the surface filling-factor) needed to produce the observed transit 
depth can be estimated if we assume a photospheric temperature to spot 
temperature ratio ($T_p/T_s$).  The star has an effective temperature of 
$T_{eff} \approx 3470$K \citep{briceno05} and M-dwarf spots can be $500-1000$K 
cooler than the photosphere \citep{bjj11}.  The optical (0.65 \micron) 
detections of the transit event have depths that range from $3\%-5\%$ 
corresponding to spot filling factors of $\sim 3.5\%-8\%$. After taking into 
account the reduced contrast with the photosphere at longer wavelengths, the 
0.58\% infrared (4.5\micron) transit depth corresponds to a filling-factor 
coverage of $\sim 1.5\%-2.5\%$ - a factor of $2 - 4$ times smaller. To explain 
the short transition duration of 0.1 in phase ($\sim 36^\circ$ in longitude), 
such a spot would need to have a viewable longitudinal extent of $\sim 36/2 
\sim 18^\circ$, and to yield a spot that covers $2\%-8\%$ of the surface, the 
viewable latitudinal extent of the spot would need to be $\sim 20^\circ - 
40^\circ$, but a spot of such latitudinal extent would likely be visible 
for more than just 1/10 of the rotation period.

All of this is not impossible for an active low-mass star, but the transit 
events would need to change shape and depth (and be flat-bottomed as is the case 
for the 2009 data).  At the same time, the spots would need to disappear and 
reappear at nearly the same stellar longitude over the course of $>3$ years.  As 
indicated in section~\ref{ephm-sub}, the transit times agree with the predicted 
ephemerides to within $\lesssim 20$ minutes or about $\lesssim 3$\% of the 
period.  The spot would need to appear, evolve, disappear, and re-appear all 
within $\lesssim 10^\circ$ of the same stellar longitudinal position as the 
event timing ephemerides have been consistent for over three years, and do this 
all within the context of a generally spotted star producing the continuous 
variability of the star.  

As in \citet{vaneyken12}, we find it difficult to model the transit events in 
a self-consistent manner with the data acquired over 3 years.  We are, however, 
continuing to pursue multi-year and multi-color light curves in an effort to 
understand better the nature of the stellar variability.

In section~\ref{ephm-sub}, we refer to the brightening event in the 2012 
December optical light curve that occurs at phase 0.8. This brightening does not 
look like a typical flare with a sharp rise and exponential decline 
\citep{walkowicz11}.  One possibility is that this brightening is the result of 
an accretion event on the star.  PTFO~8-8695 exhibits relatively strong 
H-$\alpha$ emission and it has been noted that the planetary candidate is 
near or at the Roche-limit -- particularly, if the putative orbit is 
eccentric \citep{vaneyken12}.  As a result, the planetary candidate may be 
evaporating and the ``flare'' may actually be the result of infall onto the 
stellar surface.  More detailed work on the H-$\alpha$ emission and its 
variability is the subject of another paper \citep{jk15}.  

\section{Summary}\label{summary-sec}

PTFO~8-8695b was discovered in the PTF-Orion survey for transiting exoplanets by 
\citet{vaneyken12}; PTFO~8-8695b was found to be in an 0.45 day orbit and to 
have a mass of $\lesssim 4-5 M_{Jup}$.  That discovery was followed-up by a 
prediction by \citet{barnes13} that the stellar rotation axis and the orbital 
plane of the planet should precess and that the transits should change shape and 
depth and disappear and reappear with a period of $300 - 600$ days.  The two 
papers put together a coherent picture of a Jupiter-sized planetary candidate 
orbiting a $\sim$3MYr old weak-lined T Tauri star; however, the precession model 
needed confirmation and PTFO~8-8695b remained a planetary candidate with a mass 
less than $\lesssim 3 M_{Jup}$.

We have obtained follow-up observations of PTFO~8-8695 that includes Spitzer 
4.5\micron\ light curve observations, Keck NIRSPEC radial velocity observations, 
and LCOGT optical light curve observations.  The data confirm that 
the transit events are persistent over many years and the times of the transits 
are predictable from the transit timing measurements, consistent with a steady 
period of 0.448 days.  The transit events do appear to grow more shallow, 
disappear, and reappear as predicted by the precession modeling. 

However, the precession model and observations are not in perfect agreement and 
more observations are needed to place better constraints on the model and to 
confirm the planetary nature of the candidate PTFO~8-9695b.  Towards this end, 
we are pursuing additional long-term transit observations with LCOGT and radial 
velocity monitoring that will help limit the models, the orbit, the stellar 
parameters, and help confirm or refute the planetary nature as the source of the 
observed transit events which have remained for over three years of 
observations.

\acknowledgments
DRC would like to dedicate this paper to his dad Robert A.~Ciardi (1940 -- 
2013).  Robert Ciardi had a passion for learning and knowledge and, in 
particular, for science. While not able to pursue a career in science for 
himself, he never stopped thinking and growing, and through his love and 
encouragement, DRC was able to pursue his own love of exploration and science.  
In many ways,  Robert Ciardi was more excited about this discovery than the 
authors on this paper.  He will be greatly missed. Thank you, Dad.\\  

Some of the data presented herein were obtained at the W. M. Keck Observatory, 
which is operated as a scientific partnership among the California Institute of 
Technology, the University of California and the National Aeronautics and Space 
Administration. The Observatory was made possible by the generous financial 
support of the W. M. Keck Foundation. The authors wish to recognize and 
acknowledge the significant cultural role and reverence that the summit of 
Mauna Kea has always had within the indigenous Hawaiian community. We are most 
fortunate to have the opportunity to conduct observations from this mountain.

This research has made use of the LCOGT Archive, which is operated by the 
California Institute of Technology, under contract with the Las Cumbres 
Observatory Global Telescope Network.  This work is based, in part, on 
observations made with the Spitzer Space Telescope, which is operated by the Jet 
Propulsion Laboratory, California Institute of Technology under a contract with 
NASA.  This research has made use of the NASA Exoplanet Archive, which is 
operated by the California Institute of Technology, under contract with the 
National Aeronautics and Space Administration under the Exoplanet Exploration 
Program.

Portions of this work were performed at the California Institute of Technology 
under contract with the National Aeronautics and Space Administration.

Facilities: \facility{Spitzer(IRAC)}, \facility{KeckII(NIRSPEC)}, 
\facility{LCOGT}


\end{document}